\documentclass[twocolumn]{aastex631}
\usepackage[nolist]{acronym}
\usepackage{comment}
\usepackage{upgreek}

\newcommand{\ergcm}[1]{erg~cm$^{-2}$ s$^{-1}$}
\newcommand{\g}{G288.8--6.3}

\newcommand{\HI}{{H{\sc i}}\xspace}
\newcommand{\HII}{{H{\sc ii}}\xspace}
\newcommand{\SII}{[S\,{\sc ii}]}

\newcommand{\D}{$^\circ$}

\def\HII{\hbox{H{\sc ii}}}
\def\HI{\hbox{H{\sc i}}}

\def\arcmin{\hbox{$^\prime$}}
\def\arcsec{\hbox{$^{\prime\prime}$}}

\newcommand{\ujybm}{\,$\upmu$Jy\,beam$^{-1}$}

\newcommand{\um}{\,$\upmu$m}

\begin{acronym}[AWGN]
\acro{2MASS}{Two Micron All Sky Survey}
\acro{2MASX}{Two Micron All Sky Survey Extended Source Catalogue}
\acro{30Dor}[30 Dor]{30~Doradus}
\acro{AGN}{active galactic nuclei}
\acro{ATCA}{Australia Telescope Compact Array}
\acro{ATESP}{Australia Telescope ESO Slice Project}
\acro{ATNF}{Australia Telescope National Facility}
\acro{ATOA}{Australia Telescope Online Archive}
\acro{AT20G}{Australia Telescope 20 GHz Survey}
\acro{ASKAP}{Australian Square Kilometre Array Pathfinder}
\acro{ARC}{Australian Research Council}
\acro{BETA}{Boolardy Engineering Test Array}
\acro{BL}[BL Lac]{BL Lacertae Objects}
\acro{CASS}{CSIRO Astronomy and Space Science}
\acro{CABB}{Compact Array Broadband Back-end}
\acro{CHII}[{\sc CHii}]{Compact \textsc{Hii}}
\acro{CMB}{Cosmic Microwave Background}
\acro{CSIRO}{Australian Commonwealth Scientific and Industrial Research Organisation}
\acro{CSS}{Compact Steep Spectrum}
\acro{DS9}[\textsc{DS9}]{\textsc{SAOImage DS9}}
\acro{DSA}{diffusive shock acceleration}
\acro{DSS}{Digital Sky Survey}
\acro{EM}{Electromagnetic}
\acro{EMU}{Evolutionary Map of the Universe}
\acro{eROSITA}{extended R\"Ontgen Survey with an Imaging Telescope Array}
\acro{ev}[eV]{electronvolt\acroextra{: 1 eV $\approx 1.6 \times 10^{-19}$ J}}
\acro{FITS}[\textsc{Fits}]{Flexible Image Transport System}
\acro{FRB}{Fast Radio Bursts}
\acro{FSRQ}{Flat Spectrum Radio Quasars}
\acro{FWHM}{Full Width at Half-Maximum}
\acro{GLEAM}{GaLactic Extragalactic All-sky MWA}
\acro{GPS}{Gigahertz Peak Spectrum}
\acro{HEMT}{High Electron Mobility Transistor}
\acro{HFP}[HFP]{High-Frequency Peaker}
\acro{HPBW}{Half Power Beam Width} 
\acro{HzRGs}{High Redshift Radio Galaxies}
\acro{pHFP}[pHFP]{Potential High Frequency Peaker}
\acro{HI}[H{\sc i}]{Neutral Atomic Hydrogen} 
\acro{HST}{\textit{Hubble Space Telescope}}
\acro{ICM}{intracluster medium}
\acro{IAU}{International Astronomical Union}
\acro{IFRSs}{Infrared Faint Radio Sources}
\acro{ISM}{Interstellar Medium}
\acro{IFRS}{Infrared Faint Radio Source}
\acro{IR}{Infrared}
\acro{JY}[Jy]{Jansky\acroextra{, 1 Jy = $10^{-26} \times \mathrm{W~ m}^{-2}~\mathrm{Hz}^{-1}$}} 
\acro{LLS}{Largest Linear Size}
\acro{LFAA}{Low-Frequency Aperture Array}
\acro{LMC}{Large Magellanic Cloud}
\acro{LOFAR}{Low-Frequency Array}
\acro{LSO}[LSOs]{Large Scale Objects}
\acro{MACHO}{Massive Astrophysical Compact Halo Objects}
\acro{MC}[MCs]{Magellanic Clouds}
\acro{mc}[MC]{Magellanic Cloud}    
\acro{MCELS}{Magellanic Cloud Emission Line Survey}
\acro{MW}{Milky Way}
\acro{MIRIAD}[\textsc{Miriad}]{Multichannel Image Reconstruction, Image Analysis and Display}
\acro{MIT}{Massachusetts Institute of Technology}
\acro{MOST}{Molonglo Observatory Synthesis Telescope}
\acro{MQS}{Magellanic Quasars Survey}
\acro{MRC}{Molonglo Reference Catalogue of Radio Sources}
\acro{MWA}{Murchison Widefield Array}
\acro{NRAO}{National Radio Astronomy Observatory}
\acro{NVSS}{NRAO VLA Sky Survey}
\acro{OPAL}{Online Proposal Applications \& Links}
\acrodefplural{ORC}{Odd Radio Circles}
\acro{ORC}{Odd Radio Circle}
\acro{OVV}{Optically Violent Variable Quasars}            
\acro{PAF}{Phased Array Feed}
\acro{pc}{parsec\acroextra{: 1 pc $\simeq 3.09 \times 10^{16}$ m}}
\acro{PMN}{Parkes-MIT-NRAO}
\acro{PNe}{Planetary Nebulae}
\acro{PWN}{Pulsar Wind Nebulae}
\acro{QSO}{Quasi-Stellar Object}
\acro{RA}{Right Ascension}
\acro{RFI}{Radio-Frequency Interference}
\acro{RMS}[rms]{Root Mean Squared}
\acro{SARAO}{South African Radio Astronomy Observatory}
\acro{SDSS}{Sloan Digital Sky Survey}
\acro{SED}{spectral energy distribution}
\acro{SI}[$\alpha$]{Spectral Index\acroextra{, $S \propto \nu^\alpha$}}
\acro{SKA}{Square Kilometre Array}
\acro{SMB}{Super Massive Blackholes}
\acro{SMC}{Small Magellanic Cloud}
\acro{SN}{Supernova}
\acro{SUMSS}{Sydney University Molonglo Sky Survey}
\acro{TOPCAT}[\textsc{Topcat}]{Tool for OPerations on Catalogues And Tables}
\acro{USS}{Ultra Steep Spectrum}
\acro{YSOs}{young stellar objects}
\acro{WBAC}{Wide-Band Analogue Correlator}
\acro{WIFES}[WiFeS]{Wide-Field Spectrograph}
\acro{WISE}{Wide-Field Infrared Survey Explorer}
\acro{VLA}{Very Large Array} 
\acro{VLBI}{Very Long Baseline Interferometry} 
\acro{VLSR}[\textbf{$v_{lsr}$}]{Velocity in the Line of Sight}
\acro{SMASH}{Survey of the MAgellanic Stellar History}
\acro{SNR}{Supernova Remnant}
\acrodefplural{SNR}{Supernova Remnants}
\acro{SD}{single-degenerate}
\acro{SUMSS}{Sydney University Molonglo Sky Survey}
\acro{SMBH}{Super Massive Black Hole}
\end{acronym}

\newcommand{\Sect}{Section}

\begin{document}

\title{EMU Detection of a Large and Low Surface Brightness Galactic SNR G288.8--6.3}

\author[0000-0002-4990-9288]{Miroslav D. Filipovi\'c}
\correspondingauthor{Miroslav D. Filipovi\'c}
\email{m.filipovic@westernsydney.edu.au}
\affiliation{Western Sydney University, Locked Bag 1797, Penrith South DC, NSW 2751, Australia}

\author[0000-0002-9618-2499]{Shi Dai}
\affiliation{Western Sydney University, Locked Bag 1797, Penrith South DC, NSW 2751, Australia}

\author[0000-0002-8036-4132]{Bojan Arbutina}
\affiliation{Department of Astronomy, Faculty of Mathematics, University of Belgrade, Studentski trg 16, 11000 Belgrade, Serbia}

\author[0000-0002-5119-4808]{Natasha Hurley-Walker}
\affiliation{International Centre for Radio Astronomy Research, Curtin University, Bentley, WA 6102, Australia}

\author[0000-0002-8312-6930]{Robert Brose}
\affiliation{Dublin Institute for Advanced Studies, Astronomy \& Astrophysics Section, DIAS Dunsink Observatory, Dublin, D15 XR2R, Ireland}

\author[0000-0003-1173-6964]{Werner Becker}
\affiliation{Max-Planck-Institut f\"{u}r extraterrestrische Physik, Gie{\ss}enbachstra{\ss}e 1, D-85748 Garching, Germany}
\affiliation{Max-Planck-Institut für Radioastronomie, Auf dem Hügel 69, 53121 Bonn, Germany}

\author[0000-0003-2062-5692]{Hidetoshi Sano}
\affiliation{Faculty of Engineering, Gifu University, 1-1 Yanagido, Gifu 501-1193, Japan}

\author[0000-0003-0665-0939]{Dejan Uro\v sevi\'c}
\affiliation{Department of Astronomy, Faculty of Mathematics, University of Belgrade, Studentski trg 16, 11000 Belgrade, Serbia}

\author[0000-0002-4939-734X]{T.H. Jarrett}
\affiliation{Department of Astronomy, University of Cape Town, Private Bag X3, Rondebosch 7701, South Africa}
\affiliation{Western Sydney University, Locked Bag 1797, Penrith South DC, NSW 2751, Australia}

\author[0000-0002-6097-2747]{Andrew M. Hopkins}
\affiliation{School of Mathematical and Physical Sciences, 12 Wally’s Walk, Macquarie University, NSW 2109, Australia}

\author[0000-0001-5609-7372]{Rami Z. E. Alsaberi}
\affiliation{Western Sydney University, Locked Bag 1797, Penrith South DC, NSW 2751, Australia}

\author[0000-0002-4934-7422]{R. Alsulami}
\affiliation{School of Physical Sciences, The University of Adelaide, Adelaide 5005, Australia}

\author[0000-0002-7703-0692]{Cristobal Bordiu}
\affiliation{INAF -- Osservatorio Astrofisico di Catania, via Santa Sofia 78, I-95123 Catania, Italia}

\author[0009-0003-2088-9433]{Brianna Ball}
\affiliation{Dominion Radio Astrophysical Observatory, Herzberg Astronomy and Astrophysics, National Research Council Canada, PO Box 248, Penticton BC V2A 6J9, Canada}

\author[0000-0002-3429-2481]{Filomena Bufano}
\affiliation{INAF -- Osservatorio Astrofisico di Catania, via Santa Sofia 78, I-95123 Catania, Italia}

\author[0000-0002-7239-2248]{Christopher Burger-Scheidlin}
\affiliation{Dublin Institute for Advanced Studies, Astronomy \& Astrophysics Section, DIAS Dunsink Observatory, Dublin, D15 XR2R, Ireland}
\affiliation{School of Physics, University College Dublin, Belfield, Dublin, D04 V1W8, Ireland}

\author[0000-0001-5197-1091]{Evan Crawford}
\affiliation{Western Sydney University, Locked Bag 1797, Penrith South DC, NSW 2751, Australia}

\author[0000-0001-5310-1022]{Jayanne English}
\affiliation{Department of Physics and Astronomy, University of Manitoba, Winnipeg, Manitoba R3T 2N2, Canada}

\author[0000-0002-0107-5237]{Frank Haberl}
\affiliation{Max-Planck-Institut f\"{u}r extraterrestrische Physik, Gie{\ss}enbachstra{\ss}e 1, D-85748 Garching, Germany}

\author[0000-0002-3137-473X]{Adriano Ingallinera}
\affiliation{INAF -- Osservatorio Astrofisico di Catania, via Santa Sofia 78, I-95123 Catania, Italia}

\author[0000-0002-5289-5729]{Anna D. Kapinska}
\affiliation{National Radio Astronomy Observatory, PO Box 0, Socorro, NM87801, USA}

\author[0000-0001-6872-2358]{Patrick J. Kavanagh}
\affiliation{Department of Experimental Physics, Maynooth University, Maynooth, Co. 
Kildare, Ireland}

\author[0000-0003-4351-993X]{B\"arbel S. Koribalski}
\affiliation{CSIRO Space and Astronomy, Australia Telescope National Facility, PO Box 76, Epping, NSW 1710, Australia}
\affiliation{Western Sydney University, Locked Bag 1797, Penrith South DC, NSW 2751, Australia}

\author[0000-0001-5953-0100]{Roland Kothes}
\affiliation{Dominion Radio Astrophysical Observatory, Herzberg Astronomy and Astrophysics, National Research Council Canada, PO Box 248, Penticton BC V2A 6J9, Canada}

\author[0000-0001-6109-8548]{Sanja Lazarevi\'c}
\affiliation{Western Sydney University, Locked Bag 1797, Penrith South DC, NSW 2751, Australia}
\affiliation{Astronomical Observatory, Volgina 7, 11060 Belgrade, Serbia}

\author[0000-0002-5449-6131]{Jonathan Mackey}
\affiliation{Dublin Institute for Advanced Studies, Astronomy \& Astrophysics Section, DIAS Dunsink Observatory, Dublin, D15 XR2R, Ireland}
\affiliation{School of Physics, University College Dublin, Belfield, Dublin 4, Ireland}

\author[0000-0002-9516-1581]{Gavin Rowell}
\affiliation{School of Physical Sciences, The University of Adelaide, Adelaide 5005, Australia}

\author[0000-0002-4814-958X]{Denis Leahy}
\affiliation{Department of Physics and Astronomy, University of Calgary, Calgary, Alberta, T2N 1N4, Canada}

\author[0000-0001-5126-1719]{Sara Loru}
\affiliation{INAF -- Osservatorio Astrofisico di Catania, via Santa Sofia 78, I-95123 Catania, Italia}

\author[0000-0003-1584-5930]{Peter J. Macgregor}
\affiliation{Western Sydney University, Locked Bag 1797, Penrith South DC, NSW 2751, Australia}
\affiliation{CSIRO Space and Astronomy, Australia Telescope National Facility, PO Box 76, Epping, NSW 1710, Australia}

\author[0000-0001-8534-6788]{Luciano Nicastro}
\affiliation{INAF -- Osservatorio di Astrofisica e Scienza dello Spazio di Bologna, Via Piero Gobetti 93/3, I-40129 Bologna, Italy}

\author[0000-0002-4597-1906]{Ray P. Norris}
\affiliation{Western Sydney University, Locked Bag 1797, Penrith South DC, NSW 2751, Australia}
\affiliation{CSIRO Space and Astronomy, Australia Telescope National Facility, PO Box 76, Epping, NSW 1710, Australia}

\author[0000-0001-6368-8330]{Simone Riggi}
\affiliation{INAF -- Osservatorio Astrofisico di Catania, via Santa Sofia 78, I-95123 Catania, Italia}

\author[0000-0001-5302-1866]{Manami Sasaki}
\affiliation{Dr Karl Remeis Observatory, Erlangen Centre for Astroparticle Physics, Friedrich-Alexander-Universit\"{a}t Erlangen-N\"{u}rnberg, Sternwartstra{\ss}e 7, 96049 Bamberg, Germany}

\author[0000-0002-0338-9539]{Milorad Stupar}
\affiliation{Western Sydney University, Locked Bag 1797, Penrith South DC, NSW 2751, Australia}

\author[0000-0002-1216-7831]{Corrado Trigilio}
\affiliation{INAF -- Osservatorio Astrofisico di Catania, via Santa Sofia 78, I-95123 Catania, Italia}

\author[0000-0002-6972-8388]{Grazia Umana}
\affiliation{INAF -- Osservatorio Astrofisico di Catania, via Santa Sofia 78, I-95123 Catania, Italia}

\author[0000-0001-7093-3875]{Tessa Vernstrom}
\affiliation{ICRAR, The University of Western Australia, 35 Stirling Hw, 6009 Crawley, Australia}

\author[0000-0001-9393-8863]{Branislav Vukoti\'c}
\affiliation{Astronomical Observatory, Volgina 7, 11060 Belgrade, Serbia}

\begin{abstract}

We present the serendipitous detection of a new Galactic \ac{SNR}, \g, using data from the \ac{ASKAP}-\ac{EMU} survey. Using multi-frequency analysis, we confirm this object as an evolved Galactic \ac{SNR} at high Galactic latitude with low radio surface brightness and typical \ac{SNR} spectral index of $\alpha = -0.41\pm0.12$. To determine the magnetic field strength in \ac{SNR} \g, we present the first derivation of the equipartition formulae for \acp{SNR} with spectral indices $\alpha>-0.5$. The angular size is 1\fdg8$\times$1\fdg6 (107\farcm6$\times$98\farcm4) and we estimate that its intrinsic size is $\sim$40\,pc which implies a distance of $\sim$1.3\,kpc and a position of $\sim$140\,pc above the Galactic plane. This is one of the largest angular size and closest Galactic \acp{SNR}. Given its low radio surface brightness, we suggest that it is about 13\,000 years old.

\end{abstract}

keywords{Supernova Remnants --- Radio astronomy --- \g}

\section{Introduction} 
\label{sec:intro}

The census of the Galactic \ac{SNR} population is well understood to be under-represented \citep{2013A&A...549A.107F,2021A&A...651A..86D,2023MNRAS.524.1396B}. Only $\sim$300 such objects are confirmed \citep{Green,Ferrand2012} with expectations that up to an almost order of magnitude more could exist in the Milky Way \citep{2022ApJ...940...63R}. Most of the missing Galactic \acp{SNR} are expected to be low-surface brightness or in complex regions where clear distinctions from other source types (e.g. \HII~regions) are challenging. At the same time, new, bright, small size (compact) and presumably young \acp{SNR} are not likely to be found in abundance \citep{2021Univ....7..338R}. 

With their improved capabilities and low surface-brightness sensitivity, the latest generation of radio telescopes, including \ac{ASKAP}, MeerKAT, \ac{MWA} and \ac{LOFAR}, are expected to discover many of these missing Galactic \acp{SNR}. Indeed, several recent discoveries such as the intergalactic \ac{SNR} J0624--6948 \citep{2022MNRAS.512..265F}, the Galactic \ac{SNR}s G181.1--9.5 \citep{2017A&A...597A.116K}, Hoinga \citep{2021A&A...648A..30B}, G118.4+37.0 \citep[Calvera;][]{2022A&A...667A..71A}, J1818.0--1607 \citep{2023ApJ...943...20I} and G121.1--1.9 \citep{2023MNRAS.521.5536K} confirm this expectation. These \acp{SNR} are mainly located well outside the Galactic Plane where they can preserve their typical circular \ac{SNR} shape for longer timeframes, in the presumably low-density environment while also becoming lower surface-brightness as compared to typical \acp{SNR}.

The majority of known \acp{SNR} were discovered in radio surveys with mature instruments, such as the \ac{VLA}, Parkes, the \ac{ATCA}, Effelsberg, \ac{MOST}, and the synthesis telescope at the Dominion Radio Astrophysical Observatory (DRAO ST) as part of the Canadian Galactic Plane Survey \citep[CGPS, ][]{cgps}. In addition to established \acp{SNR}, many candidates have also been identified and await confirmation. For example, \citet{1997MNRAS.287..722D} and \citet{2007Ap&SS.307..423S,2008MNRAS.390.1037S} revealed several filamentary and low-surface brightness nebulosities from the \ac{PMN} survey \citep{1994ApJS...91..111W} and the AAO/UKST H{\ensuremath{\alpha}} survey as possible candidate \acp{SNR}. 
Here we report the detection of a new Galactic \ac{SNR} which could be a representative of the missing Galactic \ac{SNR} population.

In Section~\ref{sec:eqvptr}, we first derive the equipartition formulae for determining magnetic fields in \acp{SNR} with spectral indices $\alpha>-0.5$, which will be applied to \ac{SNR} \g. Observations and data analysis are described in Seciton~\ref{sec:3}. We present and discuss our results in Section~\ref{sec:results}. Finally, the conclusions are given in Section~\ref{sec:conclusion}.

%%%%%%%%%%%%%%%%%%%%%%%%%%%%%%%% Fig. 1 %%%%%%%%%%%%%%%%%%%%%%%%%%%%%%
\begin{figure*}
    \centering\includegraphics[width=\linewidth,clip]{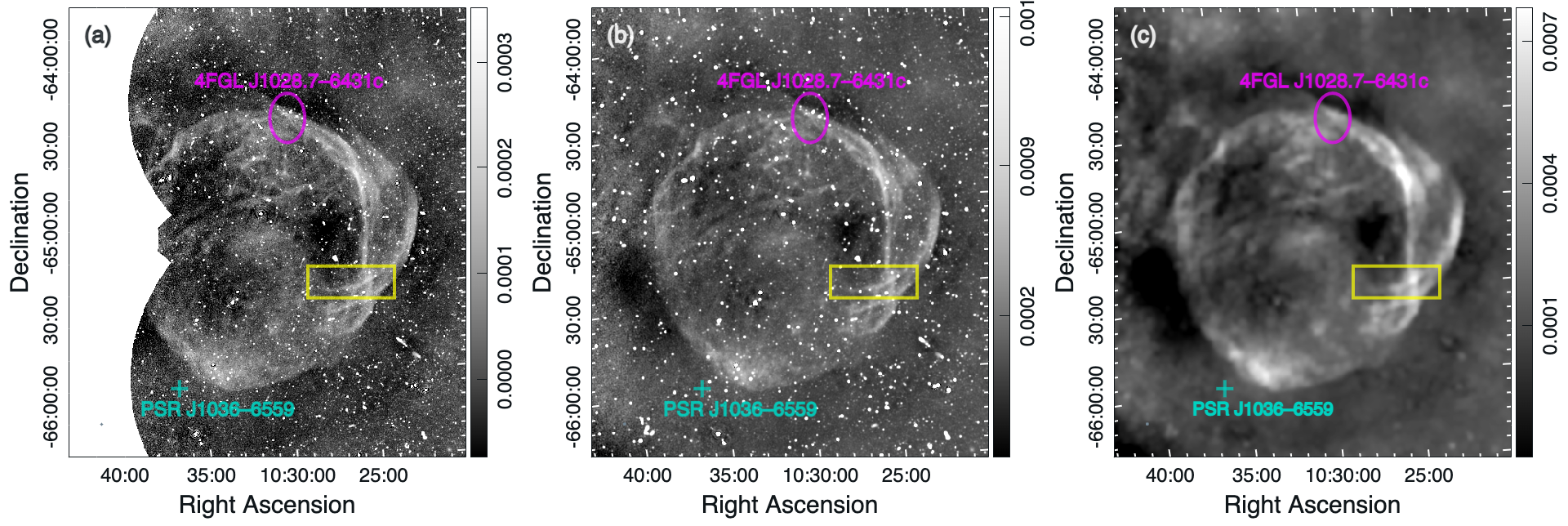}
    \caption{\ac{ASKAP} image of the Galactic \ac{SNR} \g\ at 943\,MHz. On the left (a) is the publicly released image with a synthesised beam of 15\arcsec$\times$15\arcsec\ while in the middle (b) we show an extended and re-processed image that is convolved to the resolution of 30\arcsec$\times$30\arcsec. On the right in the image (c), to enhance the diffuse emission we show image (b) that is scaled with a 90\arcsec\ box using the multi-resolution filtering method of \citet{2002PASP..114..427R}. The magenta ellipse marks the position of gamma-ray source 4FGL~J1028.7--6431c (68\% containment), while the cyan cross marks the position of nearby radio pulsar J1036--6559. The yellow rectangular box marks the area covered in Figure~\ref{fig:X}. The grey scale (in Jy\,beam$^{-1}$) is displayed on the right side, with the (a) and (b) images using a logarithmic scale and the (c) image using a linear scale. 
    }
    \label{fig:1}
\end{figure*}
%%%%%%%%%%%%%%%%%%%%%%%%%%%%%%%%%%%%%%%%%%%%%%%%%%%%%%%%%%%%%%%%

\section{Adapted Equipartition Calculation for $\alpha>-0.5$}
 \label{sec:eqvptr}

The \ac{DSA} theory in the test-particle regime predicts that for strong shocks (with compression $\sim$ 4) the particle spectrum will be a power-law with energy index  $\mathnormal{\Gamma} = 2$ \citep{1978MNRAS.182..147B}, which leads to the radio spectral index $\alpha = - \frac{\mathnormal{\Gamma}-1}{2} = -0.5$. Observed SNRs indeed have an average spectral index $\sim - 0.5$ \citep{2012SSRv..166..231R, 2014SerAJ.189...15G, 2017ApJS..230....2B,2019A&A...631A.127M,book2}, but there are remnants with $\alpha > -0.5$, which may also be the case with \g,  as we shall see in the Subsection~\ref{sec:radio}. 

Energy indices $\mathnormal{\Gamma} < 2$ can not be explained by the test-particle \ac{DSA}, but assuming that the energy index is not directly related to shock compression, we can still try to estimate the magnetic field strength for such remnants through an adapted equipartition calculation.
The synchrotron flux density can be written as \citep{2017POBeo..97....1A}
\begin{equation}
S _\nu = \frac{4\pi}{3} f \theta ^3 d\ c_5 K_e (B \sin \Theta )^{(\mathnormal{\Gamma} +1)/2}
\Big(\frac{\nu}{2c_1}\Big)^{(1-\mathnormal{\Gamma} )/2},
\end{equation}
where $c_1$ and $c_5$ are defined in \citet{1970ranp.book.....P}, $B$ is the magnetic field strength, $\Theta$ the pitch angle, $K_e$ the constant in the  power-law energy distributions of cosmic ray (CR) electrons $N(E)=K_e E^{-\mathnormal{\Gamma}}$, $\nu$ is the frequency, $f$ is the volume filling factor, $\theta$ the angular
radius and $d$ is the distance. We shall use averaged sine of the pitch angle, 
$ 
\frac{\sqrt{\pi}}{2}\frac{\Gamma(\frac{\mathnormal{\Gamma} +5}{4})}{\Gamma(\frac{\mathnormal{\Gamma} +7}{4})}$. 
From the above equation, using spectral index $\alpha = - \frac{\mathnormal{\Gamma}-1}{2}$ we can write
\begin{equation}
 K_e B ^{\alpha +1} = J,
\end{equation}
where $J$ is a function that can be calculated from observable quantities. To estimate $B$, we need to know some relation between $K_e$ and $B$, which is the basis for the minimum-energy or equipartition calculation \citep{1970ranp.book.....P}. In \ac{SNR}s, the equipartition between CRs energy and magnetic field energy does not seem to be reached. Still, the magnetic field energy density may be comparable to the energy density of CR electrons \citep{2018ApJ...855...59U}.

CR energy density for a power-law particle distribution of electrons $f(p)= k p^{-\mathnormal{\Gamma} -2}$ \citep{2012ApJ...746...79A, 2013ApJ...777...31A, 2017POBeo..97....1A} can be written as 
\begin{eqnarray}
\epsilon _\mathrm{CR} &= & \int _{p_{\mathrm{inj}}} ^{p_\infty} 4\pi k
p^{-\mathnormal{\Gamma}} (\sqrt{p^2c^2+m_e^2c^4}-m_ec^2){d}p \nonumber
\\
&= & 4\pi k c (m_e c)^{2-\mathnormal{\Gamma}} \int _{\frac{p_{\mathrm{inj}}}{m_e c}}
^{\frac{p_\infty}{m_e c}} x^{-\mathnormal{\Gamma}} (\sqrt{x^2 + 1}-1){d}x, \ \ \
x=\frac{p}{m_ec} \nonumber
\\
 &= & K_e (m_ec^2)^{2-\mathnormal{\Gamma}} I_1,
\end{eqnarray}
where $K_e = 4 \pi k c^{\mathnormal{\Gamma} -1}$.
Injection momentum can be written as
$
 p_\mathrm{inj} = \xi  p_\mathrm{th} = \xi \sqrt{2 m_e kT_e},
$
where $\xi$ is injection parameter related to the injection efficiency \citep{2005MNRAS.361..907B}
\begin{equation}
 \eta = \frac{4}{\sqrt{\pi}} \frac{1}{\Gamma -1} \xi ^3 e^{-\xi ^2}.
\end{equation}
For $T_e$, we can provisionally assume $T_e \sim 0.4 T_p$ \citep[for $v_s \sim 600$\,km\,s$^{-1}$][]{2013SSRv..178..633G}, where $T_p$ can be calculated from Rankine-Hugoniot jump conditions.

Assuming synchrotron losses, the maximum electron energy \citep{2007A&A...465..695Z} is 
\begin{equation}
E_\infty = p_\infty c
 = \frac{1}{1 + \sqrt{\kappa}} \frac{\sigma-1}{3\sigma} \frac{m_e^2 c^3 v_s}{\sqrt{\frac{2}{27}e^3B}},
\end{equation}
where $\sigma$ shock compression, $\kappa = B_0/B = 1/\sqrt{(1+2\sigma ^2)/{3}}$, $B_0$ ambient magnetic field. 
If $\epsilon _\mathrm{CR} \approx \epsilon _B = \frac{1}{8\pi} B^2$ then
\begin{equation}
K_e = \frac{(m_e c)^{\mathnormal{\Gamma}-2}}{8 \pi I_1} B^2,
\end{equation}
and the magnetic field estimate is
\begin{equation}
B^{\alpha +3} = 8 \pi (m_e c)^{2-\mathnormal{\Gamma}} I_1 J.
\label{Eq7}
\end{equation}
If we had used proton equipartition, then
\begin{equation}
K_p = \frac{(m_p c)^{\mathnormal{\Gamma}-2}}{8 \pi I_1} B^2,
\end{equation}
and with $p_\mathrm{inj} = \xi \sqrt{2 m_p kT_p}$,  $E_\infty 
 = \frac{3}{8} \frac{v_s}{c} e B R$ and $K_{ep} = \Big(\frac{m_e}{m_p}\frac{T_e}{T_p}\Big)^\alpha$ \citep{2021APh...12702546A}, 
the magnetic field would be
\begin{equation}
B^{\alpha +3} = 8 \pi (m_p c)^{2-\mathnormal{\Gamma}} I_1 J/K_{ep}.
\label{Eq9}
\end{equation}

In Subsection~\ref{sec:radio} we will apply Eqs. (\ref{Eq7}) and (\ref{Eq9}) to estimate magnetic field strength in \g.

\section{Observations and Data}
\label{sec:3}

As a part of the large scale \ac{ASKAP}-\ac{EMU} project (AS201), this part of the radio sky was observed in January 2023 with a complete set of 36 \ac{ASKAP} antennas at the central frequency of 943.4\,MHz and bandwidth of 288~MHz. All data can be found in the CSIRO \ac{ASKAP} Science Data Archive (CASDA\footnote{\url{https://research.csiro.au/casda}}). The observations containing this object are the tile EMU\_1003$-$64 corresponding to \ac{ASKAP} scheduling block SB46974. The data were processed using the ASKAPsoft pipeline, including multi-frequency synthesis imaging, multi-scale clean and self-calibration \citep{askapsoft_2019ascl.soft12003G}. The resulting 943~MHz \ac{EMU} image has a sensitivity of $\sigma$=25\ujybm\ and a synthesised beam of 15\arcsec$\times$15\arcsec\ (Figure~\ref{fig:1}a). 

A small portion of the eastern side of \g\ is at the edge of the \ac{EMU} observing block (tile) (see Figure~\ref{fig:1}a). The \ac{EMU} project is surveying the full southern sky, enabling this serendipitous detection of \g. Currently, only Stokes $I$ and $V$ images are available.

In order to extend the view further towards the east, we re-imaged \g\ to a lower primary beam cutoff. We used visibilities for beams 5, 10, 11, 16, 17, 22, 23 and 29 of SB46974 and re-imaged with the same continuum settings of the above described \ac{ASKAP} pipeline (Figure~\ref{fig:1}b). The beams were then convolved to a common circular beam of 30\arcsec$\times$30\arcsec\ and mosaiced together using the holography beams from the SB metadata. We use a lower weight cutoff of 0.01, instead of the default 0.20 to extend the coverage of the mosaic to the east. This cutoff may introduce errors as the holography measurements only just sample the primary field of view \citep{aces11}, and may not fully characterise the edge beams imaged here. We measured the image \ac{RMS} of $\sigma$=31\ujybm.

Finally, we used the multiresolution filtering method of \citet{2002PASP..114..427R} to enhance the diffuse emission of \g\ as seen in Figure~\ref{fig:1}c.

\begin{figure*}
    \centering\includegraphics[width=\linewidth]{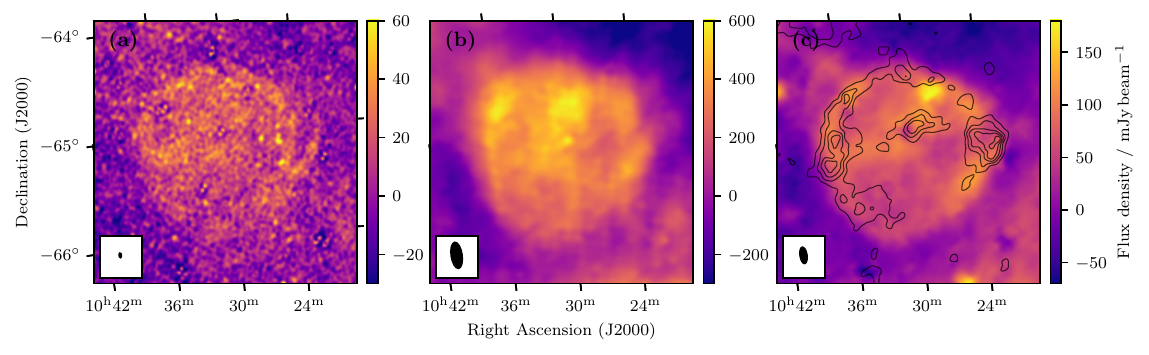}    
    \caption{$2\fdg5\times2\fdg5$ region surrounding \g{} at \ac{MWA} 200\,MHz (a), CHIPASS 1400\,MHz (b), and S-PASS 2300\,MHz (c) after source and background subtraction as described in \Sect~\ref{sec:radio}. The contours on the third panel show the total linear polarisation measured in S-PASS, in steps of 4$\times$ the local RMS noise, 38\,$\upmu$Jy\,beam$^{-1}$. The full-width-half-maximum of the point spread function is shown as a solid black ellipse in the lower left of each panel.}
    \label{fig:radio_images}
\end{figure*}

We examined several large-area radio surveys to search for \g\ signatures and to derive the flux density as a function of frequency. Only surveys with sensitivity to scales at least that of its diameter (1\fdg7) can be used so that it is not resolved out. At the lowest frequency, we used data taken by the \ac{MWA} \citep[][]{2013PASA...30....7T,2018PASA...35...33W} for the GaLactic and All-sky \ac{MWA} \citep[GLEAM;][]{2015PASA...32...25W} survey, at the edge of the 170--231\,MHz source-finding mosaics generated by \cite{2017MNRAS.464.1146H} (see Figure~\ref{fig:radio_images}a).

At 1400\,MHz, we used the continuum map of the \HI\ Parkes All-Sky Survey \citep[CHIPASS\footnote{\url{https://www.atnf.csiro.au/people/mcalabre/CHIPASS/index.html}};][]{2014PASA...31....7C}, 
a radio sky survey covering Dec~$<+25\degr$ at $14\farcm4$ resolution (see Figure~\ref{fig:radio_images}b). At 2300\,MHz, we used the S-Band Polarization All Sky Survey \citep[S-PASS;][]{2019MNRAS.489.2330C}, a 2300-MHz survey of polarised radio emission covering the southern sky (Dec~$<-1\degr$) at $8\farcm9$ resolution (see Figure~\ref{fig:radio_images}c).

To remove the contaminating point sources from the data, we followed the methods previously detailed by \citet{2021A&A...648A..30B} and \citet{2022MNRAS.510.2920A}. We performed source-finding on the \ac{MWA} and EMU images, using \textsc{Aegean}\footnote{\url{https://github.com/PaulHancock/Aegean}} \citep{2012MNRAS.422.1812H,2018PASA...35...11H} and its companion tool the Background and Noise Estimator (\textsc{BANE}). We subtracted the \ac{MWA} 200-MHz point sources from the image using a further ancillary tool \textsc{AeRes}. For CHIPASS and S-PASS, we scaled the flux densities of the sources detected in the EMU image by a spectral index (defined by $S \propto \nu^{\alpha}$, where $S$ is flux density, $\nu$ is the frequency and $\alpha$ is the spectral index) reasonable at these flux densities ($\alpha=-0.83$), convolved the model to their lower resolutions and performed subtraction.

%%%%%%%%%%%%%%%%%%%%%%%%%%%%%%%% Fig. 2 %%%%%%%%%%%%%%%%%%%%%%%%%%%%%%
%%%%%%%%%%%%%%%%%%%%%%%%%%%%%%%%%%%%%%%%%%%%%%%%%%%%%%%%%%%%%%%%

Finally, we used several other large-area sky surveys such as \ac{PMN} survey (Section~\ref{sec:radio}), HI4PI (Section~\ref{sec:HI}), SuperCOSMOS (Section~\ref{sec:halpha}), \ac{WISE} (Section~\ref{sec:IR}), \ac{eROSITA} All Sky Survey (eRASS; Section~\ref{sec:xray}) and Fermi (Section~\ref{sec:gama}).

\section{Results and Discussion} 
\label{sec:results}

We have serendipitously found a large-scale object in our new \ac{ASKAP}-\ac{EMU} survey \citep{2021PASA...38...46N} which we classify as the new Galactic \ac{SNR} \g\ (Figure~\ref{fig:1}). Our detection and classification of this Galactic \ac{SNR} is primarily based on the object morphology, size and multi-frequency appearance of \g\ and the method is described in \citet[][Section 2.4]{2019PASA...36...48H}, \citet[][]{2022MNRAS.512..265F} and \citet{2023MNRAS.518.2574B}. We note that \citet{1997MNRAS.287..722D} and \citet{2008MNRAS.390.1037S} identified this object as a possible \ac{SNR} candidate but wrongly named G288.7--6.3 because they couldn't define \g\ exact central position and extent.

\subsection{SNR \g\ Radio-continuum Properties}
 \label{sec:radio}
 
We used the Minkowski tensor analysis tool BANANA\footnote{\url{https://github.com/ccollischon/banana}} \citep{2021A&A...653A..16C} to determine the centre of \g\ to be at RA(J2000)\,=\,10$^{\rm h}$30$^{\rm m}$22\fs3 and Dec(J2000)\,=\,--65\D12\arcmin46\farcs5, with a size of 1\fdg8$\times$1\fdg6 (107\farcm6$\times$98\farcm4) at a position angle of 30\D. 

\begin{figure}  
    \centering\includegraphics[width=\linewidth]{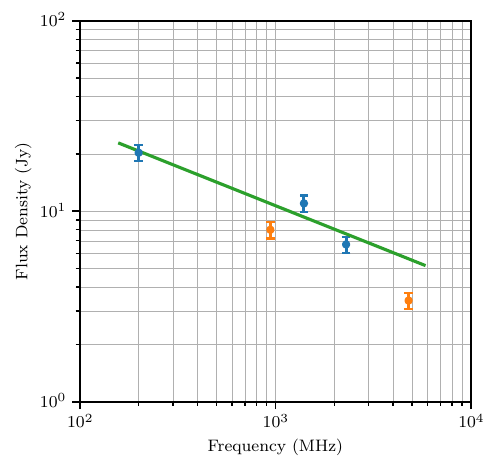}
    \caption{Radio flux densities as a function of frequency. A power-law fit based on flux density measurements (blue points) from \ac{MWA}, CHIPASS and S-PASS surveys with $S_\mathrm{1\,GHz} = 10.92\pm0.06$\,Jy and $\alpha = -0.41\pm0.12$ is shown with a green line. Two orange points are flux density measurements from \ac{ASKAP} and \ac{PMN} surveys and they are not included in the spectral index estimate.}
    \label{fig:sed}
\end{figure}

To remove gradients from the large-scale Galactic diffuse emission in \ac{MWA}, CHIPASS and S-PASS images, we fit a 2-D plane to the region around \g\ using \textsc{polygon-flux}\footnote{\href{https://github.com/nhurleywalker/polygon-flux}{https://github.com/nhurleywalker/polygon-flux}} \citep{2019PASA...36...48H}. The source-subtracted, backgrounded data are shown in Figure~\ref{fig:radio_images}. This tool was also used to obtain total radio flux densities at these three frequencies (200, 1400 and 2300~MHz) which produced a spectral index of $\alpha = -0.41\pm0.12$. Errors were assigned as 20, 10 and 10\,\% respectively. The data and a power-law fit are shown in Figure~\ref{fig:sed}.

%%%%%%%%%%%%%%%%%%%%%%%%%%%%%%%% Fig. 3 %%%%%%%%%%%%%%%%%%%%%%%%%%%%%%
%%%%%%%%%%%%%%%%%%%%%%%%%%%%%%%%%%%%%%%%%%%%%%%%%%%%%%%%%%%%%%%%

\ac{SNR} \g\ is also detected in the \ac{PMN} survey at 4850\,MHz, although the median filter applied to the data makes it impossible to obtain reliable flux densities for large objects. After a similar subtraction of point sources, we estimate lower flux density limits of $S_{\rm PMN}$=3.4\,Jy and $S_{\rm ASKAP}$=8.0\,Jy. This indicates a missing flux density of $>50\%$ at both frequencies and we did not use these two values for the spectral index estimate. Even if we included these two flux density estimates, the spectral index would be only fractionally steeper ($\alpha=-0.51\pm0.12$) but still within the range of the above estimate of $\alpha=-0.41\pm0.12$.

This spectral index of $\alpha = -0.41\pm0.12$ is somewhat flatter than that of average shell type \acp{SNR} but within the observed range of --0.55$\pm$0.20 for the Galaxy and range of nearby galaxy \acp{SNR} \citep{2012SSRv..166..231R, 2014SerAJ.189...15G, 2017ApJS..230....2B,2019A&A...631A.127M,book2}. We also note an enhanced extended radio emission in the centre of \g\ which could indicate the presence of a \ac{PWN}. However, we couldn't independently confirm this emission's nature as it could also come from the \g\ shell.

S-PASS also enables an examination of the polarisation features; the shell of \g\ shows up clearly in total linear polarisation with absolute intensities of $\sim$600\,\ujybm{} and fractionally about 1\,\% (contours in panel (c) of Figure~\ref{fig:radio_images}). An extended feature of similar brightness north of the shell centre may be a polarised filament viewed in projection or potentially confusion from the surrounding point sources. Future polarisation measurements with Polarization Sky Survey of the Universe's Magnetism \citep[POSSUM; ][]{2010AAS...21547013G} will enable better measurement of the \g\ polarisation features. 

Most \acp{SNR} are characterised by their position in the brightness--to--diameter ($\Sigma$--D) parameter space. Based on the above-measured flux densities and the angular size of \g, we estimate a radio surface brightness at 1~GHz of $\Sigma=1.4\times10^{-22}$~W~m$^{-2}$~Hz$^{-1}$~sr$^{-1}$ (assuming the emission to be spread smoothly over the \ac{SNR} boundaries defined above). This value  places \g\ in the bottom area of the \ac{SNR} $\Sigma-D$ diagram 

\citep[for a sample of 110 known Galactic \acp{SNR} used as calibrators, see][]{2019SerAJ.199...23S}
and indicates an intrinsic diameter of 40\,pc ($\pm$9\,pc), determined by the analysis of the probability density function in the $\Sigma-D$ plane and the corresponding orthogonal fitting procedure \citep[]{2019SerAJ.199...23S}. Coupled with the above measured angular size, this indicates a 1.3\,kpc ($\pm$0.4\,kpc) distance which is among the 10 closest Galactic \acp{SNR}. Given this distance range and the Galactic latitude of --6\fdg3, \g\ is 140\,pc away from the Galactic plane, suggesting that it is still inside the thin disk.

The intrinsic size of $\sim$40\,pc is close to the median size of all known \acp{SNR} in our Galaxy \citep{Green}. At one end, the largest would be the Monoceros loop (G205.5+0.5) with a diameter of $\sim$128\,pc \citep{2020ApJ...891..137Z}. At the other end, the smallest would be $\sim$140 year young \ac{SNR} G1.9+0.3 with a diameter of only 3.9\,pc \citep{2020MNRAS.492.2606L}. A similar range of \ac{SNR} diameters can be found in the \ac{LMC} \citep{2017ApJS..230....2B,2021MNRAS.500.2336Y,2023MNRAS.518.2574B} where \ac{SNR} MCSNR\,0450--70 is the largest (102$\times75\pm1$\,pc) \citep{2009SerAJ.179...55C} and SN1987A would be the smallest. 
There are only $\sim$12 known Galactic \acp{SNR} with a larger angular size listed in the catalogue of \citet{Green}. The most similar (in angular size) are Galactic \acp{SNR} G89.0+4.7 (HB21) \citep{2011A&A...529A.159G} and G7.5--1.7 \citep{2019PASA...36...45H,2019PASA...36...48H} but both of these \acp{SNR} are much brighter radio sources. 

\begin{figure}
    \centering\includegraphics[width=1\linewidth,clip]{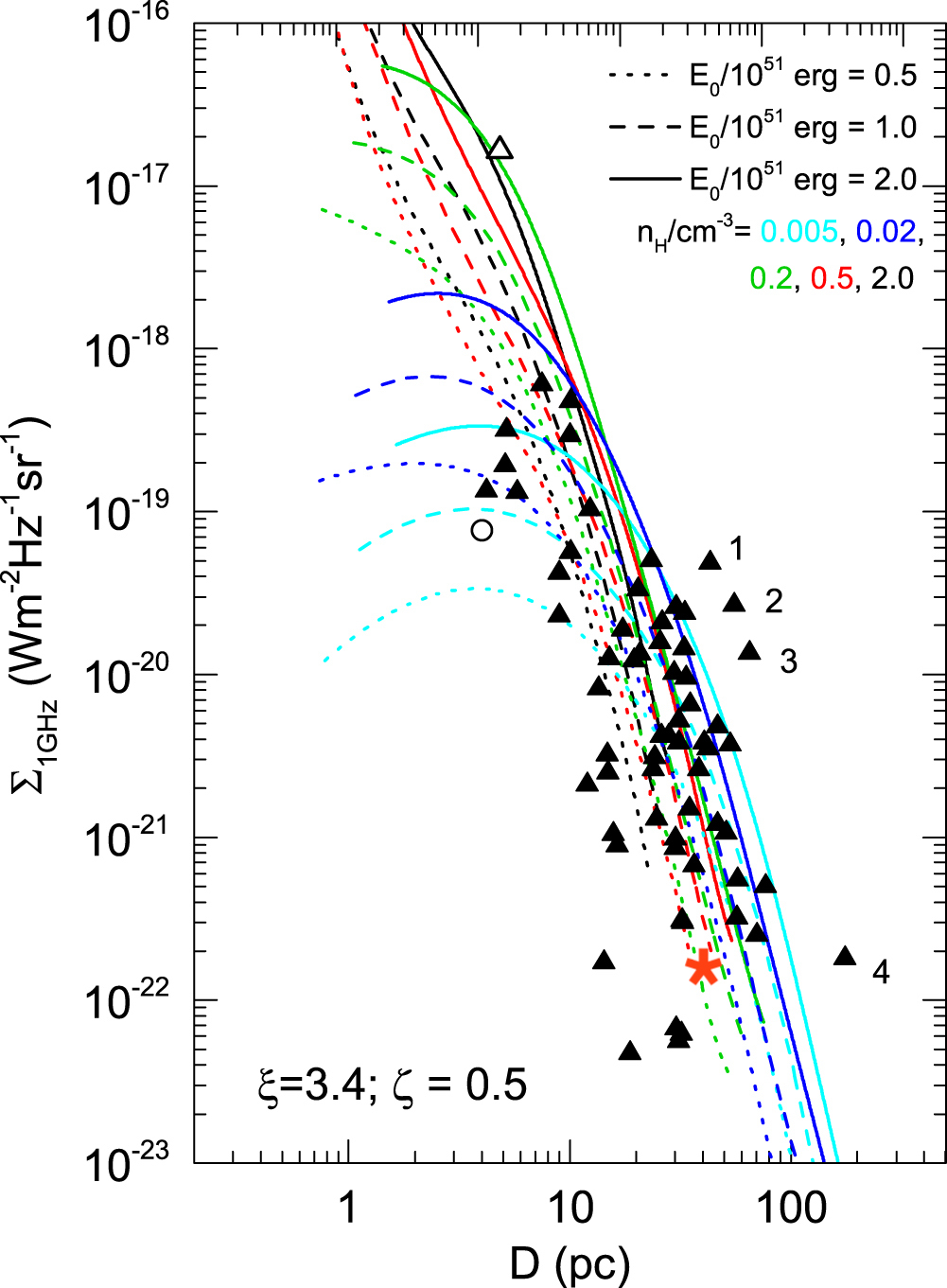}
    \caption{Radio surface brightness–to–diameter diagram for \ac{SNR}s at a frequency of 1\,GHz (black triangles), obtained from numerical simulations \citep[][their Fig.~3]{2018ApJ...852...84P}. Our newly detected \ac{SNR} \g\ is marked with a red star in the bottom right corner while Cassiopeia~A is shown with an open triangle. The open circle represents the youngest Galactic \ac{SNR}, G1.9+0.3 \citep{2020MNRAS.492.2606L}. Numbers represent the following \acp{SNR}: (1) CTB~37A, (2) Kes~97, (3) CTB~37B and (4) G65.1+0.6.}
    \label{fig:evol}
\end{figure}

\g\ has somewhat lower surface brightness (Figure~\ref{fig:evol}) than the majority of known Magellanic Clouds and Galactic \acp{SNR} with established distance \citep{2019SerAJ.199...23S}. Given the low radio surface brightness of \g, we suggest it is in an evolved Sedov stage and {about} 13\,000 years old. Assuming an ambient density of 0.17\,cm$^{-3}$ \citep{2006A&A...459..113M} and \ac{SN} energy of $E_0$ = 10$^{51}$\,erg we can estimate the shock velocity to be $v_s \sim 600$\,km\,s$^{-1}$ and the Sedov age about $\sim 10\,000$ years. Taking into account the preceding free expansion phase, a better description of shell dynamics is given by a formula \citep{2012ApJ...751...65F, 2017ApJS..230....2B}
\begin{equation}
\frac{1}{2} v_s^2= \frac{1}{2} \bigg( \frac{dR}{dt} \bigg)^2 = \frac{k_1E_0
}{k_2{M}_0+4\pi R^3 \rho _0 /3} ,
\end{equation}
where constant $k_1$ is determined so that the evolution of a shell for large radii $R$ tend toward Sedov's solution, while $k_2$ is obtained from initial conditions. 
For a rough estimate of \ac{SN} ejecta mass $M_0 \sim 5~M_\odot$ and initial velocity $v_0 = \sqrt{\frac{2k_1 E_0}{k_2 M_0}} \approx 5\,000$\,km\,s$^{-1}$, by integrating the last equation we get an age estimate of 13\,000 years. {Since} the estimates of the ambient density range from 0.11 to 0.24\,cm$^{-3}$ \citep{2003ApJ...587..278W} for an assumed distance of $1.3\,$kpc and $z$=140\,pc, {this uncertainty in density alone would give an age range $\sim$ 11\,000--15\,000 years. However, it is difficult to put exact age limits, given the uncertainty in other parameters.}

To estimate the magnetic field strength in \g\  we use the adapted equipartition calculation described previously. By applying  Eq. (\ref{Eq7}), taking for the filling factor $f=0.1$, compression $\sigma \approx 4$ and $\xi \approx 4$ (for non-modified strong shocks) and using the observable and calculated quantities of \g \ (including the 1\,GHz flux density of 10.92\,Jy) we find $B=7.7\,\upmu$G in the ambient field of 2.3\,$\upmu$G. On the other hand, if we apply Eq. (\ref{Eq9}), $B$ would be $41.7\,\upmu$G in the ambient field of 12.6\,$\upmu$G.

At the age of 13\,000 years, \g\ is among a couple of dozen known historical Galactic \acp{SNR}. One could compare its possible intrinsic brightness at the time of explosion to the well-studied SN\,1054 (Crab nebulae). With the assumption that they are of similar type of \ac{SN} explosion and relying on Chinese historical reports about the SN\,1054 brightness (as bright as Venus or $m\sim -4.5$ at a distance of 2\,kpc) at the time of the explosion \citep{10541,10542,10543}, we estimate that \g\ \ac{SN} explosion would be $m\sim -5.4$ magnitudes. This would indicate that the \g\ explosion was likely to have been seen even during the daytime in the southern hemisphere.

%%%%%%%%%%%%%%%%%%%%%%%%%%%%%%%% Fig. 4 %%%%%%%%%%%%%%%%%%%%%%%%%%%%%%
%%%%%%%%%%%%%%%%%%%%%%%%%%%%%%%%%%%%%%%%%%%%%%%%%%%%%%%%%%%%%%%%

There is no known radio pulsar or any circularly polarised compact object (with $V/I>0.5$\%) within the borders of \g. Just outside the southeast edge of \g, we find the known pulsar J1036--6559 (Figure~\ref{fig:1}) with a spin period of 0.5335\,s and dispersion measure of 158.36\,cm$^{-3}$\,pc~\citep{2012MNRAS.427.1052B}. The pulsar has a characteristic age of $\sim$6\,Myr and a surface magnetic field of $\sim$8$\times10^{11}$\,G, which is typical for normal pulsars~\citep{2005AJ....129.1993M}. 
According to the YMW16 electron density model~\citep{2017ApJ...835...29Y}, the distance to PSR~J1036--6559 is $\sim$2\,kpc.
If we assume that the two are connected, for PSR~J1036--6559 to travel from the centre of \g\ to its present position ($\sim$22\,pc; assuming motion to be perpendicular to the line of sight) over the age of 13\,000~years it would require a transverse velocity of $\sim$1650\,km\,s$^{-1}$. This somewhat large velocity for pulsars would leave a prominent trail (the bow-shock; exp. ``mouse'' \citep{2002ApJ...579L..25C}, DEM\,S5 \citep{2019MNRAS.486.2507A} or ``potoroo'' (Lazarevi\'c~et.~al. in prep.)) that should be observed either in our \ac{ASKAP} or \ac{eROSITA} images. Therefore, either \ac{SNR} \g\ is significantly older ($>$100\,000\,years) or much smaller (and closer) than $D$=40\,pc (1.3\,kpc). It is more plausible to conclude that there is no direct connection between PSR~J1036--6559 and \g.

\subsection{G288.8--6.3 H$\alpha$ Emission}
 \label{sec:halpha}

\begin{figure*}
    \centering\includegraphics[width=\linewidth]{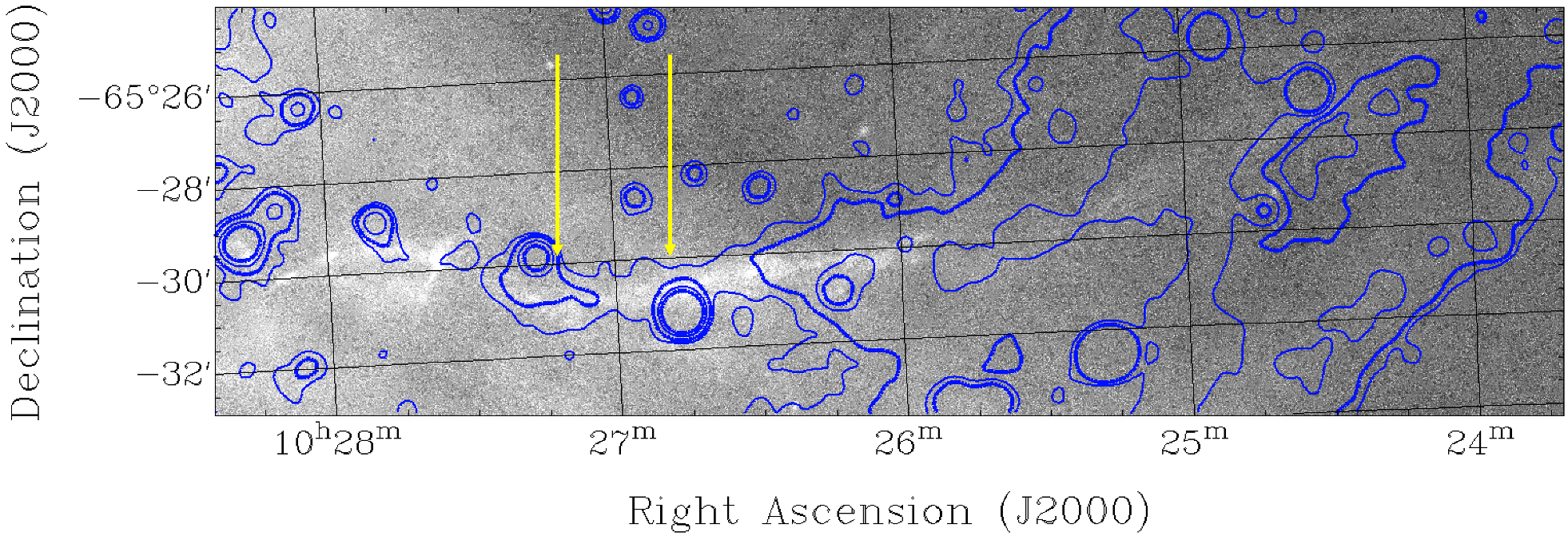}    
    \caption{H$\alpha$ SuperCOSMOS grey scale (background subtracted) image overlaid with \ac{ASKAP} contours (0.1, 0.2, 0.5 and 1~mJy\,beam$^{-1}$) of the Galactic \ac{SNR} \g. Two yellow arrows indicate possible H$\alpha$ filament likely corresponding to radio emission. The area covering this image is marked in Figure~\ref{fig:1} as the yellow rectangular box.}
    \label{fig:X}
\end{figure*}

Our initial search for signatures at other frequencies showed no signs of emission that could be directly linked to \g\ apart from three weak H$\alpha$ emission regions in the AAO/UKST H$\alpha$ SuperCOSMOS\footnote{\url{http://www-wfau.roe.ac.uk/sss/halpha/hapixel.html}} survey \citep[][]{2005MNRAS.362..689P}.
In Figure~\ref{fig:X} we show the H$\alpha$ feature which is marked with yellow arrows in the image. The zoomed frame clearly shows the arc feature overlapping with the radio contours showing likely associations. However, due to numerous point sources such as stars and galaxies in the image field, identifying large-scale features can be challenging. \citet[][see their Fig.~4]{2008MNRAS.390.1037S} report another H$\alpha$ feature (in the north west rim of \g) with typical \ac{SNR} optical spectral characteristics and the ratio \SII/H$\alpha$ of 0.54 that is associated with \g. This further supports that \g\ is Galactic \ac{SNR}.

%%%%%%%%%%%%%%%%%%%%%%%%%%%%%%%% Fig. 5 %%%%%%%%%%%%%%%%%%%%%%%%%%%%%%
%%%%%%%%%%%%%%%%%%%%%%%%%%%%%%%%%%%%%%%%%%%%%%%%%%%%%%%%%%%%%%%%

Finally, we found a foreground open cluster that overlaps the \ac{SNR} on the sky. It has an estimated age $\sim$50\,Myr and it is only $\sim$140\,pc away, with one B0 star \citep[$\theta$ Carinae; see e.g. Fig.~1 in][]{2022AJ....163..278N}. This young cluster, which still has at least one star massive enough that it should explode as a core-collapse \ac{SN}, could be home or somehow associated with \g\ if it was closer. However, the surface brightness of \g\ is so faint that it must be a physically large and middle-aged \ac{SNR}, therefore cannot be this close to Earth. One could speculate that this object could be a wind bubble or some other type of wind-driven nebula, but we find this to be an unlikely scenario because there is not an obvious star near the centre that could be driving a wind.

\subsection{H{\sc i} Study of G288.8--6.3}
 \label{sec:HI}

\begin{figure*}
    \centering\includegraphics[width=\linewidth,clip]{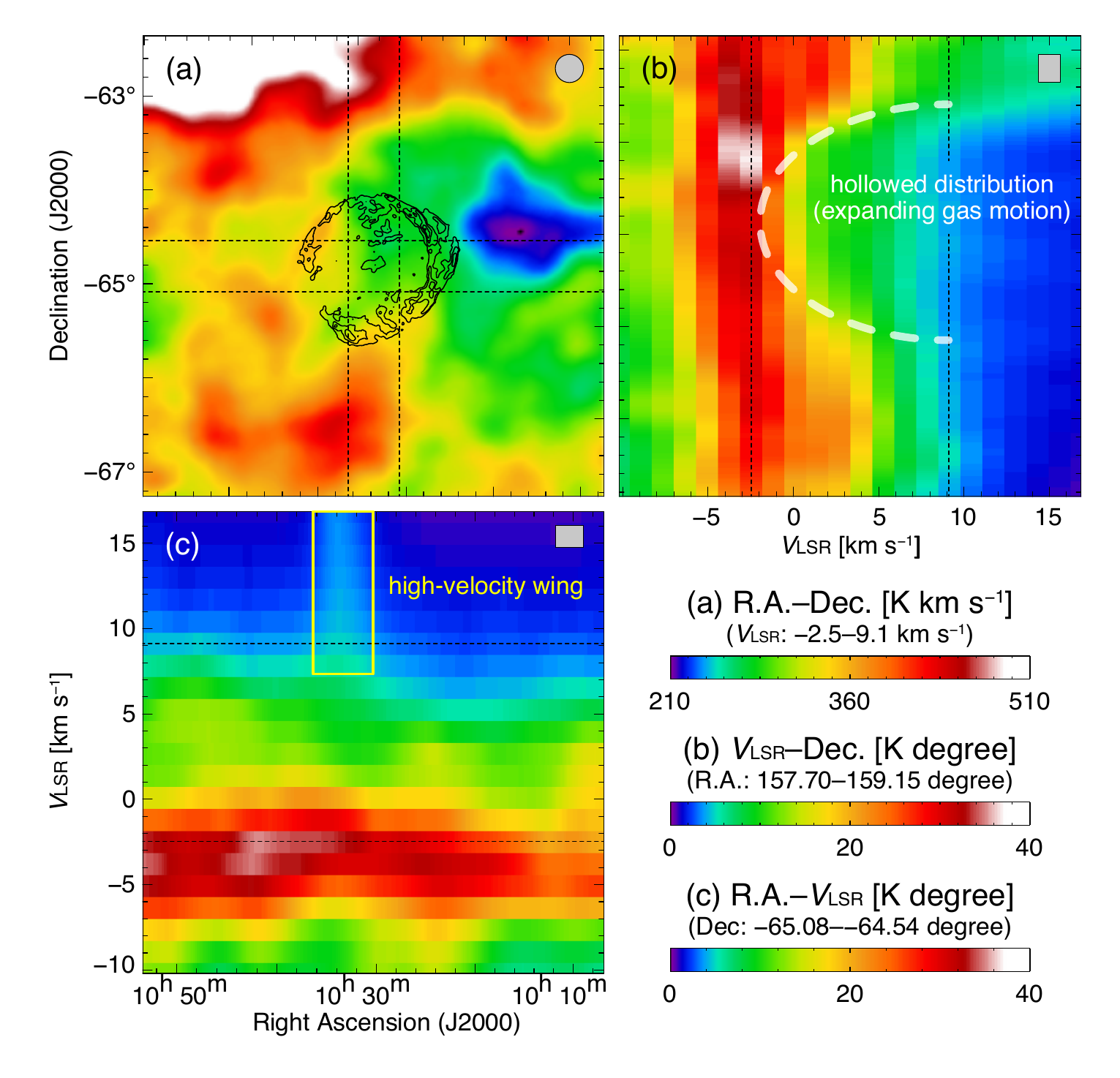}
    \caption{(a) Intensity distribution of \HI\ obtained from HI4PI \citep{2016A&A...594A.116H}. The integration velocity range is from $-2.5$ to 9.1\,km\,s$^{-1}$. The superposed contours indicate the intensity of the radio continuum. (b)--(c) Position--velocity (p--v) diagrams of \HI. The \HI\ intensity is integrated from 157\fdg70 to 159\fdg15 in Right Ascension for (b) and from $-65\fdg08$ to $-64\fdg54$ in Declination for (c). The white dashed curve and yellow solid box in the p--v diagrams indicate candidates of an \HI\ expanding shell and a high-velocity wing (see the text).}
    \label{fig:3}
\end{figure*}

In order to reveal the physical relation between the \ac{SNR} and its surroundings, we analysed the archival \HI\ data taken from HI4PI which has a modest angular resolution of $16'$ \citep{2016A&A...594A.116H}. Figure~\ref{fig:3}(a) shows the velocity-integrated intensity map of \HI\ towards the \g. We found a cavity-like distribution of \HI\ in the velocity range from $-2.5$ to 9.1\,km\,s$^{-1}$. The northeast and southern radio shell boundaries show a good spatial correspondence with \HI\ clouds as shown in yellow to orange colour, suggesting the shock-cloud interaction occurred. On the other hand, the western radio shell with a blowout structure can be understood as the shocks travelling into the low-density space as shown in green to blue colour. 

We also found other possible evidence for the shock-cloud interaction in p--v diagrams of \HI. Figure~\ref{fig:3}(b) shows a hollowed distribution of \HI\ in the $V_\mathrm{LSR}$--Dec diagram, indicating the presence of an expanding \HI\ shell formed by supernova shocks and/or strong winds from the progenitor. Although the R.A.--$V_\mathrm{LSR}$ diagram does not show such an expanding shell possibly due to the inhomogeneous gas distribution of \HI, we found a high-velocity component within the radio shell (see the yellow box in Figure~\ref{fig:3}c). In \acp{SNR}, such a high-velocity component is thought to be formed by shock acceleration and provides an alternative support for the shock--cloud interaction \citep[e.g.,][]{1990ApJ...364..178K}. 

Note that we also found another possible counterpart of \HI\ with a different velocity range in terms of spatial distributions of \HI\ clouds and the radio shell which we could not rule out because of the modest angular resolution of the HI4PI data. Nonetheless, the kinematic distance to the \ac{SNR} is expected to be small (few kpc) if the \HI\ clouds at a velocity of $\sim$0\,km\,s$^{-1}$ are physically associated with the \ac{SNR}. Further high-resolution \HI\ observations are needed to test this possibility.

%%%%%%%%%%%%%%%%%%%%%%%%%%%%%%%% Fig. 6 %%%%%%%%%%%%%%%%%%%%%%%%%%%%%%
%%%%%%%%%%%%%%%%%%%%%%%%%%%%%%%%%%%%%%%%%%%%%%%%%%%%%%%%%%%%%%%%

\subsection{WISE view on SNR \g}
 \label{sec:IR}

We create a large 3$\times$3~degrees mosaic of the \g\ \ac{SNR} and its local environment as traced in the mid-infrared by the \ac{WISE} bands of W1 (3.4\,\um), W2 (4.6\,\um), W3 (12\,\um) and W4 (23\,\um). The mosaics were created using the \ac{WISE} WXSC pipeline \citep{2012AJ....144...68J} with native resolution (6\arcsec\ in W1) and supersampled with 1\arcsec\ pixels. As described in \citet{2013AJ....145....6J,2019ApJS..245...25J}, the \ac{WISE} bands were uniquely designed to sample both stellar emission (W1 and W2), and \ac{ISM} star-formation-excited gas/dust emission (W3 and W4). As expected, the stellar density is extremely high at these low Galactic latitudes, reaching the confusion limit of the \ac{WISE} imaging resolution.

Inspection of bespoke WISE mid-infrared mosaics of the field did not yield any obvious spatial alignments with the radio continuum source. This rules out any strong dust or shock-excitation features emitted in the mid-infrared window (3-25\,\um) from the \ac{SNR} candidate.

\subsection{X-ray Observations and Data Analysis of G288.8--6.3}
 \label{sec:xray} 

The X-ray data we report here were taken during the first five \ac{eROSITA} all-sky surveys \citep{Predehl2021, Sunyaev2021}. \g\ was mapped in a total of 210 telescope passages in the following date intervals: January 8–16 and July 9–18, 2020; January 7–13 and July 15–21, 2021; January 12–18, 2022. The resulting averaged, vignetting corrected exposure time is 1607\,s ($\sim$320\,s per sky survey). The data we used in our analysis were processed by the eSASS (\ac{eROSITA} Standard Analysis Software) pipeline \citep{Brunner2022} and have the processing number $\#020$. For the data analysis, we used eSASS version 211214 (released in December~$21^{\rm st}$~2021)\footnote{\url{https://erosita.mpe.mpg.de}}. Within the eSASS pipeline, X-ray data of the eRASS sky are divided into 4700 partly overlapping sky tiles of $3\fdg6 \times 3\fdg6$ each. These are numbered using six digits, three for RA and three for Dec, representing the sky tile centre position in degrees. The majority of \g\ falls into the eRASS sky tile numbered 156156, whereas the surrounding sky tiles 155153, 161153 and 163156 are required for complete coverage of \g.

\begin{figure*}
\centering\includegraphics[width=\textwidth]{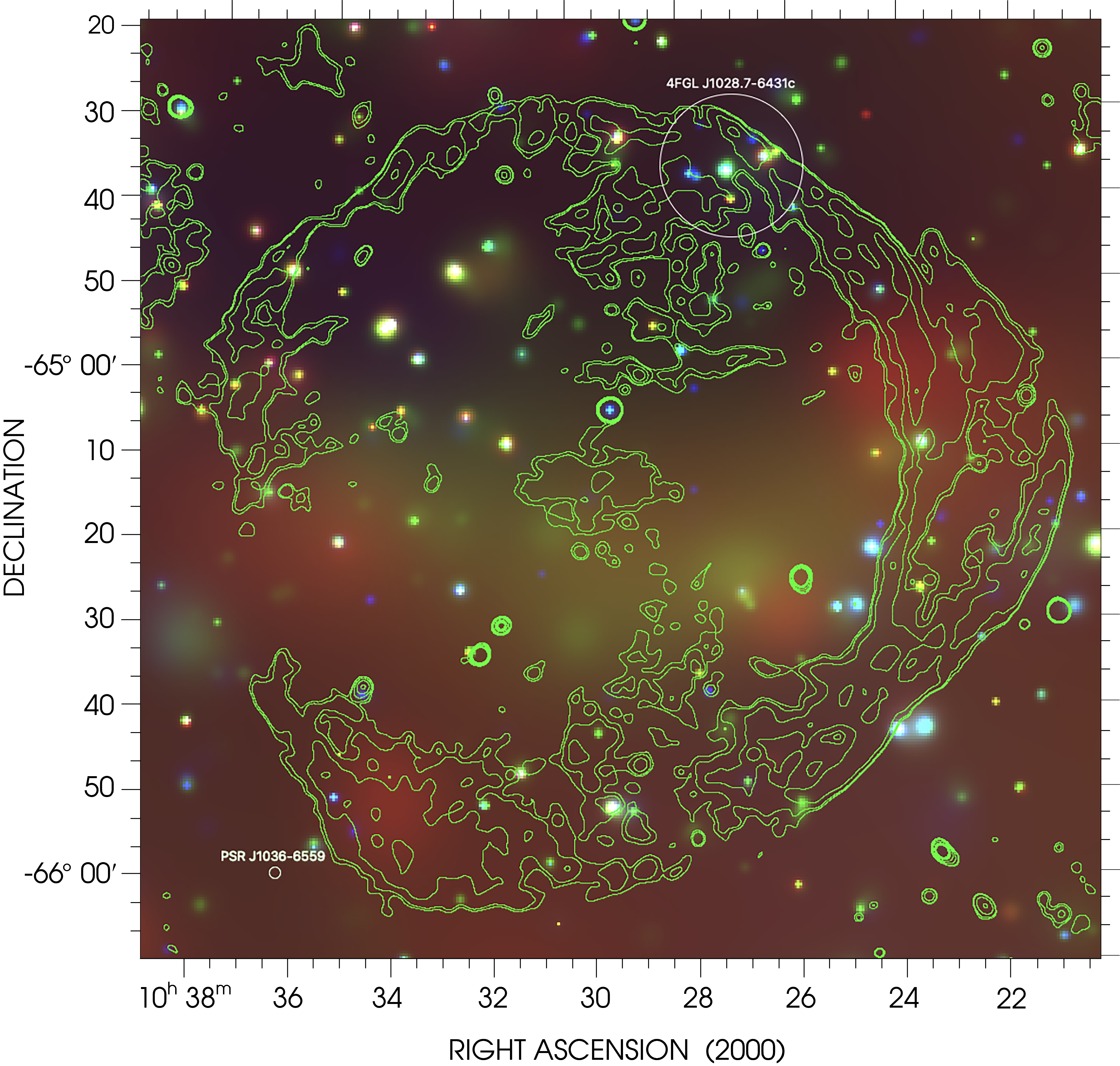}
 \caption{RGB image of \g\ as seen in the \ac{eROSITA} all-sky surveys eRASS:5. Photons to produce the image were colour coded according to their energy (red for energies 0.2--0.7\,keV, green for 0.7--1.2\,keV, blue for 1.2--2.4\,keV). An adaptive kernel smoothing algorithm was applied to the images in each energy band. Radio contour lines (green: 0.1, 0.2, 0.5 and 1\,mJy\,beam$^{-1}$) are overlaid to mark the outline of \g. The position of the neighbouring radio pulsar PSR~J1036--6559, as well as of the Fermi source 4FG~J1028.7--6431c, are shown.}
\label{fig:Fig7}
\end{figure*}

Figure~\ref{fig:Fig7} depicts an RGB image of \g\ which has been colour coded according to the energy of the detected X-ray photons. To produce the RGB image, we first created images for the three energy bands 0.2--0.7\,keV, 0.7--1.2\,keV, and 1.2--2.4\,keV, respectively. \ac{eROSITA}'s FOV averaged angular resolution during survey mode is $26\arcsec$. The spatial binning in Figure~\ref{fig:Fig7} was set to $20\arcsec$ in order to compensate for a small resolution reduction by the smoothing process. Data from all seven telescopes were used, as we did not notice a significant impact of the light leak in TM5 and TM7 \cite[][]{Predehl2021}. In order to enhance the visibility of diffuse emission in the RGB image whilst leaving point sources unsmoothed to the greatest possible extent, we applied the adaptive kernel smoothing algorithm of \cite{2006MNRAS.368...65E} with a Gaussian kernel of $3.0~\sigma$. As can be seen from Figure~\ref{fig:Fig7} there is faint diffuse X-ray emission filling the inner  parts of \g\ in the soft (0.2--0.7\,keV) and medium (0.7--1.2\,keV) energy band, partly overlapping with the radio contour lines of the remnant in the west and south-east. Whether this emission can be associated with \g\ or whether it is part of an unrelated diffuse background structure is unclear. The low photon statistics do not allow us to perform a detailed spectral analysis.

%%%%%%%%%%%%%%%%%%%%%%%%%%  Figure 7 %%%%%%%%%%%%%%%%%%%%%
%%%%%%%%%%%%%%%%%%%%%%%%%%%%%%%%%%%%%%%%%%%%%%%%%%%%%%

\subsection{Gamma-ray Observations of G288.8--6.3}
 \label{sec:gama}

\acp{SNR} are firmly established as gamma-ray emitters. Usually, bright gamma-ray emission is expected from \acp{SNR} in dense environments where pp-interactions create $\pi_0$s that consequently decay into gamma-rays. Also, many bright TeV \acp{SNR} are quite young (less than a few kyr) and may have a significant inverse-Compton component from electrons. Older \acp{SNR} (beyond the synchrotron cooling time for electrons) however could have significant $\pi^{0}$-decay components.
However, the brightest known \acp{SNR} in TeV gamma-rays are all expanding into very low-density environments, similar to \g, where instead, high peak luminosities might be reached via inverse-Compton scattering \citep{2020A&A...634A..59B}. 

The closest known gamma-ray source to \g\ is 4FGL\,J1028.7--6431c from the Fermi-LAT catalogue of extended sources. 4FGL\,J1028.7--6431c is located 0\fdg7 away from the centre of \g\ and hence roughly overlapping with the northern shell of the \ac{SNR} (see Figure~\ref{fig:1}). The gamma-ray source has a detection significance of $5.6\,\sigma$ and the spectrum is described by a log parabola of the form
\begin{equation}
    F(E) = F_0\left(\frac{E}{E_0}\right)^{- \alpha - \beta \log(E/E_0)}\text{ , } 
\end{equation}
with $F_0 = (1.5\pm0.3)\times10^{-12}\,$cm$^{-2}$MeV$^{-1}$s$^{-1}$, $\alpha=2.6\pm0.3$ and $\beta=0.4\pm0.2$ and a pivot energy of $E_0=720\,$MeV. The source is slightly elongated with a 68\% containment radius of roughly 0.12$^\circ$. Given that the gamma-ray emission is overlapping with a bright part of the radio shell, an association with \g\ is likely but an association with PMN~J1028--6441 was claimed earlier \citep{2022ApJS..260...53A}. 

The integrated photon flux of \g\ is $F_{1-100\text{GeV}}=(2.6\pm0.6)\times 10^{-10}\,$s$^{-1}$cm$^2$. This is somewhat lower than the fluxes reported for other high-latitude extended Fermi sources such as FHES\,J1741.6--3917 ($(47.5 \pm 4.6)\times 10^{-10}\,$s$^{-1}$cm$^2$) and FHES\,J1208.7--5229 ($(9.6 \pm 1.6)\times 10^{-10}\,$s$^{-1}$cm$^2$) that are potentially associated with the well-evolved \acp{SNR} G351.0--5.4 and G296.5+9.7 \citep{2018ApJS..237...32A}. However, given that the emission from 4FGL\,J1028.7--6431c is only originating from a part of the shell, emission from the rest of the remnant might be missed in the automated analysis. Further, \cite{2021A&A...654A.139B} showed that the emission morphology of evolved \acp{SNR} can extend well beyond the radio-emitting shell and becomes centre-filled for IC-dominated gamma-ray emitters due to the escape of accelerated particles. Hence, a dedicated analysis of Fermi-LAT data of the region is encouraged.

\subsection{Evolutionary Status}
 \label{sec:evstat}

To estimate the evolutionary status of this newly observed \ac{SNR}, we apply the method presented in \citet{2020NatAs...4..910U,2022PASP..134f1001U}. Using \citet[][]{2018ApJ...852...84P} $\Sigma-D$ evolutionary tracks, we can conclude that \g\ is an evolutionary advanced and low surface brightness \ac{SNR} in the late Sedov or radiative phase of evolution (Figure~\ref{fig:evol}). It evolves in a medium-density interstellar environment with densities $\sim$0.2\,cm$^{-3}$. The energy of the \ac{SN} explosion can be assumed to be the canonical 10$^{51}$\,erg. Additionally, {\g}'s somewhat flatter spectral index ($\alpha = -0.41$) indicates possible Fermi~2 acceleration in addition to the \ac{DSA} which can be characteristic of older \acp{SNR} \citep{2013Ap&SS.346....3O,2014Ap&SS.354..541U}. The slight influence for appearing of flatter spectral indices in evolutionary evolved \acp{SNR} can be provided by the inclusion of thermal bremsstrahlung emission in addition to synchrotron emission, especially for \acp{SNR} that evolve in dense environments which is not the case for \g, but some fraction of the spectral index flattening can be explained by this kind of influence \citep{2007ApJ...655L..41U,2012ApJ...756...61O}. The next important issue for the determination of the evolutionary stage for \g\ is from associated magnetic field values. In Section~\ref{sec:eqvptr}, we presented the derivation of the equipartition method for spectral index $\alpha>-0.5$ (adapted equipartition calculation). Both obtained values for the electron (7.7\,$\upmu$G, ambient field 2.3\,$\upmu$G) and proton (41.7\,$\upmu$G, ambient field 12.6\,$\upmu$G) equipartitions lead to the unique conclusion: shock-compression of the magnetic field (lower than a factor of 4) in this \ac{SNR} is sufficient to produce large enough magnetic fields to explain the synchrotron emission of the \acp{SNR}. Further amplification of the magnetic field by the non-linear \ac{DSA} effects is not required to explain the emission.

\section{Conclusions}
\label{sec:conclusion}

We discovered a nearby ($\sim$1.3\,kpc), large angular size (1\fdg8$\times$1\fdg6) and low surface brightness galactic \ac{SNR} \g\ in the \ac{ASKAP}-\ac{EMU} survey. Its radio spectral index ($\alpha = -0.41\pm0.12$) is typical for an evolutionary advanced \ac{SNR} population in the late adiabatic or in the radiative phases of evolution which evolves in a medium-density interstellar environment. We estimated its intrinsic size to be 40\,pc, and its low radio surface brightness (at 1~GHz of $\Sigma=1.4\times10^{-22}$~W~m$^{-2}$~Hz$^{-1}$~sr$^{-1}$) suggests that it is about 13\,000 years old. Based on the intrinsic size and the distance, it is located $\sim$140\,pc above the Galactic plane. 

We derived equipartition formulae for determining \g\ magnetic fields with spectral indices $\alpha>-0.5$ and estimated the magnetic field strength between 7.7\,$\upmu$G and 41.7\,$\upmu$G in the ambient field of 2.3\,$\upmu$G to 12.6\,$\upmu$G, respectively. 

From our \HI\ study, we found a cavity-like distribution and possible evidence for the shock-cloud interaction. We detect some of the \g\ filaments in H$\alpha$ SuperCOSMOS survey and  possible association with a gamma-ray source from the Fermi-LAT catalogue of extended sources -- 4FGL\,J1028.7--6431c. \ac{WISE} and \ac{eROSITA} X-ray observations also point to some tantalising emissions that might be associated with \g. 

Future in-depth polarimetric and multifrequency studies will enhance our knowledge of this large angular size Galactic object.
Finally, given \ac{ASKAP}'s sensitivity to low surface brightness emission, we anticipate many more similar discoveries as the \ac{EMU} survey progresses.

\section{Acknowledgments}
This scientific work uses data obtained from Inyarrimanha Ilgari Bundara / the Murchison Radio-astronomy Observatory. We acknowledge the Wajarri Yamaji People as the Traditional Owners and native title holders of the Observatory site. The \ac{CSIRO}’s \ac{ASKAP} radio telescope is part of the Australia Telescope National Facility\footnote{\url{https://ror.org/05qajvd42}}. Operation of \ac{ASKAP} is funded by the Australian Government with support from the National Collaborative Research Infrastructure Strategy. \ac{ASKAP} uses the resources of the Pawsey Supercomputing Research Centre. Establishment of \ac{ASKAP}, Inyarrimanha Ilgari Bundara, the \ac{CSIRO} Murchison Radio-astronomy Observatory and the Pawsey Supercomputing Research Centre are initiatives of the Australian Government, with support from the Government of Western Australia and the Science and Industry Endowment Fund. 
\ac{eROSITA} is the soft X-ray instrument aboard SRG, a joint Russian-German science mission supported by the Russian Space Agency (Roskosmos), in the interests of the Russian Academy of Sciences represented by its Space Research Institute (IKI), and the Deutsches Zentrum für Luft- und Raumfahrt (DLR). The SRG spacecraft was built by Lavochkin Association (NPOL) and its subcontractors and is operated by NPOL with support from IKI and the Max Planck Institute for Extraterrestrial Physics (MPE). The development and construction of the \ac{eROSITA} X-ray instrument was led by MPE, with contributions from the Dr.~Karl Remeis Observatory Bamberg \& ECAP (FAU Erlangen-N\"urnberg), the University of Hamburg Observatory, the Leibniz Institute for Astrophysics Potsdam (AIP), and the Institute for Astronomy and Astrophysics of the University of T\"ubingen, with the support of DLR and the Max Planck Society. The Argelander Institute for Astronomy of the University of Bonn and the Ludwig Maximilians Universit\"at Munich also participated in the science preparation for \ac{eROSITA}. The \ac{eROSITA} data shown here were processed using the eSASS/NRTA software system developed by the German \ac{eROSITA} consortium.
MDF, GR and SL acknowledge \ac{ARC} funding through grant DP200100784. 
N.H.-W. is the recipient of \ac{ARC} Future Fellowship project number FT190100231.
SD is the recipient of an \ac{ARC} Discovery Early Career Award (DE210101738) funded by the Australian Government.
HS acknowledges funding from JSPS KAKENHI Grant Number 21H01136.
DU and BA acknowledge the financial support provided by the Ministry of Science, Technological Development and Innovation of the Republic of Serbia through the contract 451-03-47/2023-01/200104, and for support through the joint project of the Serbian Academy of Sciences and Arts and Bulgarian Academy of Sciences on the detection of Galactic and extragalactic \acp{SNR} and \HII\ regions. 
RB acknowledges funding from the Irish Research Council under the Government of Ireland Postdoctoral Fellowship program. 
JM acknowledges support from a Royal Society-Science Foundation Ireland University Research Fellowship (20/RS-URF-R/3712).
CBS acknowledges support from a Royal Society Research Fellows Enhancement Award 2021 (22/RS-EA/3810).
BV is funded by the Ministry of Education, Science and Technological Development of the Republic of Serbia through contract number 451-03-47/2023-01/200002.
PJK acknowledges support from the Science Foundation Ireland/Irish Research Council Pathway programme under Grant Number 21/PATH-S/9360.
We thank an anonymous referee for comments and suggestions that greatly improved our paper.

\vspace{5mm}
\facilities{\ac{ASKAP}, \ac{MWA}, Parkes, \ac{WISE}, \ac{eROSITA}, Fermi-LAT}

\software{ASKAPsoft, AEGEAN (BANE) }

\bibliography{G288}{}
\bibliographystyle{aasjournal}

\end{document}